\documentstyle[seceq,epsf]{ptptex}

\preprintnumber{} 
\title{
Post-Newtonian Expansion of Gravitational Waves 
from a Particle in Circular Orbits around a Rotating Black Hole\\
{{\em\it -- Effects of Black Hole Absorption  --}} }
\author{Hideyuki {\sc Tagoshi}, 
Shuhei {\sc Mano}$^{\dag}$ and Eiichi {\sc Takasugi}$^\dag$}
\inst{
National Astronomical Observatory, 
Mitaka 181 \\
$^\dag$Department of Physics, 
Osaka University, Toyonaka 560  
}

\abst{
When a particle moves around a Kerr black hole, it radiates gravitational waves.
Some of these waves are absorbed by the black hole.
We calculate such absorption of gravitational waves 
induced by a particle of mass $\mu$ in a circular orbit on an equatorial plane 
around a Kerr black hole of mass $M$. 
We assume that the velocity of the particle $v$ 
is much smaller than the speed of light $c$ 
and calculate the energy absorption rate analytically. 
We adopt an analytic technique for the Teukolsky equation developed by 
Mano, Suzuki and Takasugi. 
We obtain the energy absorption rate to $O((v/c)^8)$ compared to the 
lowest order. We find that the black hole absorption occurs at 
$O((v/c)^5)$ beyond the Newtonian-quadrapole luminosity 
at infinity in the case when the black hole is rotating, which is 
$O((v/c)^3)$ lower than the non-rotating case. 
Using the energy absorption rate, we investigate its effects 
on the orbital evolution of coalescing compact binaries. 
}
\begin{document}

\maketitle

\section{Introduction}
Among the possible sources of gravitational waves, 
coalescing compact binaries are 
the most promising candidates for detection by near-future, 
ground based laser interferometric 
detectors such as LIGO, 
VIRGO, GEO600 and TAMA. 
When a signal of gravitational waves is detected, 
we will attempt to extract parameters for binaries, such as 
masses and spins etc., from inspiral wave forms 
using the matched filtering technique. 
In this method, parameters for binaries are determined 
by cross-correlating the noisy signal from detectors 
with theoretical templates. 
If the signal and the templates lose phase with each other by 
one cycle over $\sim 10^3 -10^4$ cycles 
as the waves sweep through the interferometer band, their 
cross correlation will be significantly reduced. 
This means that, in order to extract information optimally, 
we need to make theoretical templates 
which are accurate to better than one cycle during an entire sweep 
through the interferometer's band. \cite{ref:three}

The standard method to calculate inspiraling wave forms 
from coalescing binaries 
is the post-Newtonian expansion of the Einstein equations, 
in which the orbital velocity $v$ of the binaries 
are assumed to be small compared to the speed of light. 
Since, for coalescing binaries, the orbital velocity is not 
so small when the frequency of gravitational waves is in LIGO/VIRGO
band, it is necessary to carry the post-Newtonian expansion 
up to extremely high order in $v$. A post-Newtonian wave generation 
formalism which can handle the high order 
calculation has been developed by Blanchet, Damour 
and Iyer.\cite{ref:BandD,ref:DI} 
Based on this formalism, calculations have been carried out 
up to post$^{5/2}$-Newtonian order, or $O(v^5)$ beyond the leading 
order quadrupole formula.\cite{ref:blanchet} 
Another formalism has been developed up to $O(v^4)$ 
by Will and Wiseman. \cite{ref:WW} 
This formalism is based on that of Epstein and Wagoner.\cite{ref:EWW} 

Another post-Newtonian expansion techniques based on black hole 
perturbation formalism has also been developed. 
In this analysis, 
one considers gravitational waves from a particle of mass $\mu$ 
orbiting a black hole of mass $M$ when $\mu \ll M$. 
Although this method is restricted to the case $\mu\ll M$, 
one can calculate very high order post-Newtonian corrections to 
gravitational waves using a relatively simple algorithm 
in contrast with the standard post-Newtonian analysis. 
The gravitational radiation at infinity was calculated 
using this method to $O(v^8)$ 
by Tagoshi, Shibata, Tanaka and Sasaki \cite{ref:TSTS} 
in a Kerr black hole case, and to $O(v^{11})$ by 
Tanaka, Tagoshi and Sasaki \cite{ref:TTS} in a Schwarzschild black 
hole case. 

Despite these works, a study of gravitational waves which are absorbed by 
a black hole has been developed very little in the context of the post-Newtonian 
approximation. The post-Newtonian approximation of 
the absorption of gravitational waves into the black hole horizon 
was first calculated by Poisson and 
Sasaki in the case when a test particle is in a circular orbit around a 
Schwarzschild black hole.\cite{ref:PoSa} 
In this case, since the effect of the black hole absorption appears at 
$O(v^8)$, this effect is negligible for the orbital evolution of  
coalescing compact binaries in the above laser interferometer's band. 
However, as we shall see in this paper, 
black hole absorption appears at $O(v^5)$ when a 
black hole is rotating. Thus it is important to investigate 
the effect of black hole absorption to the orbital evolution of 
coalescing compact binaries when at least
one of the stars is a rotating black hole. 

In order to calculate the post-Newtonian expansion of ingoing 
gravitational waves into a Schwarzschild black hole, 
Poisson and Sasaki\cite{ref:PoSa} used two types of representations of 
a solution of the homogeneous Teukolsky equation. 
One is expressed in terms of spherical Bessel functions 
which can be used at large radius, 
and the other is expressed in terms of a hypergeometric function which 
can be used near the horizon. Then the two types of expressions are matched at 
some region where both formulas can be applied. 
They obtained formulas for a solution of the Teukolsky equation which can be 
used to calculate ingoing gravitational waves to $O(v^{13})$, although
they presented formulas for ingoing waves only to $O(v^8)$. 

On the other hand, another analytic technique for the Teukolsky equation
was found by Mano, Suzuki and Takasugi.\cite{ref:MST} Since this method is 
very powerful for the calculation of the post-Newtonian expansion of 
the Teukolsky equation, we adopt this method in this paper. 

This paper is organized as follows. 
In \S 2, we develop the Teukolsky equation. In \S 3, 
first we describe the analytic techniques for the Teukolsky equation and 
then we solve the Teukolsky equation analytically assuming the frequency of the
wave is small. In \S 4, the energy absorption rate is calculated 
to $O(v^8)$ compared to the lowest order. 
In \S 5, we consider the effect of black hole absorption on the 
orbital evolution of coalescing compact binaries. 
Section 6 is devoted to summary and discussion.

Throughout this paper we use units in which $c=G=1$.

\newpage
\section{The Teukolsky formalism }
In the Teukolsky formalism,\cite{ref:Teu} 
the gravitational perturbations of the Kerr black 
hole are described by a Newman-Penrose quantity 
$\psi_4=-C_{\alpha\beta\gamma\delta}
n^{\alpha}\bar{m}^\beta n^\gamma \bar{m}^\delta$. Here 
$C_{\alpha\beta\gamma\delta}$
is the Weyl tensor, $n^\alpha=((r^2+a^2),-\Delta,0,a)/(2\Sigma)$, 
$m^\alpha=(ia\sin\theta,0,1,i/\sin\theta)/(\sqrt{2}(r+ia\cos\theta))$. 
We use the Boyer-Lindquist 
coordinates $(t,r,\theta,\phi)$ and 
$\Sigma=r^2+a^2\cos^2\theta$, $\Delta=r^2-2Mr+a^2$, 
where $a$ is the spin parameter of a Kerr black hole. 
We decompose $\psi_4$ into Fourier-harmonic components according to
\begin{equation}
(r-i a \cos\theta)^4\psi_4=
\sum_{\ell m}\int d\omega 
e^{-i\omega t+i m \varphi} \ _{-2}S_{\ell m}(\theta)
R_{\ell m\omega}(r). 
\end{equation}
The radial function $R_{\ell m\omega}$ and the angular function 
$_{s}S_{\ell m}(\theta)$ satisfy the Teukolsky 
equations with $s=-2$ as 
\begin{equation}
\Delta^2{d\over dr}\left({1\over \Delta}{dR_{\ell m\omega}\over dr}
\right)
-V(r) R_{\ell m\omega}=T_{\ell m\omega}, 
\end{equation}
\begin{eqnarray}
& &\Bigl[{1 \over \sin\theta}{d \over d\theta}
 \Bigl\{\sin\theta {d \over d\theta} \Bigr\}
-a^2\omega^2\sin^2\theta
 -{(m-2\cos\theta)^2 \over \sin^2\theta}
\nonumber\\
& &~~~~~~~~~~~~~~~
+4a\omega\cos\theta-2+2ma\omega+\lambda\Bigr] 
{}_{-2}S_{\ell m}=0. \label{eq:spheroid}
\end{eqnarray}
The potential $V(r)$ is given by
\begin{equation}
V(r)=-{K^2+4i(r-M)K \over \Delta}+8i\omega r+\lambda,
\end{equation} 
where $K=(r^2+a^2)\omega-ma$ and $\lambda$ is the eigenvalue 
of $_{-2}S_{\ell m}^{a\omega}$. 
The angular function $_{s}S_{\ell m}(\theta)$ is 
the spin-weighted spheroidal harmonic which can be normalized as 
\begin{equation}
\int_0^{\pi} |_{s}S_{\ell m}|^2 \sin\theta d\theta=1.
\label{eq:snorm}
\end{equation}
The source term $T_{\ell m\omega}$ is given 
in a paper TSTS. 

We define two kinds of homogeneous solutions of the radial 
Teukolsky equation: 
\begin{eqnarray}
&&R^{{\rm in}}_{\ell m\omega} \rightarrow\cases{
C_{\ell m\omega}\Delta^2  e^{-ikr^*}\,&
for $r\rightarrow r_+$ \cr
r^3 A_{\ell m\omega}^{{\rm  out}}e^{i\omega r^*}+
r^{-1}A_{\ell m\omega}^{{\rm in}} e^{-i\omega r^*}\,&
for $r\rightarrow +\infty,$ \cr}
\label{eq:Rin} 
\\
&&R^{{\rm up}}_{\ell m\omega} \rightarrow\cases{ 
B_{\ell m\omega}^{{\rm  out}}e^{ik r^*}+ 
\Delta^2 B_{\ell m\omega}^{{\rm in}} e^{-ik r^*}\,& 
for $r\rightarrow r_+$ \cr
D_{\ell m\omega}r^3  e^{i\omega r^*}\,& 
for $r\rightarrow +\infty$, \cr
} 
\label{eq:Rup} 
\end{eqnarray} 
where $k=\omega-ma/2Mr_+$, and 
$r^*$ is the tortoise coordinate defined by 
\begin{eqnarray}
r^{*}&=&\int {dr^*\over dr}dr \nonumber\\
&=&r+{2Mr_{+}\over {r_{+}-r_{-}}}\ln{{r-r_{+}}\over 2M}
-{2Mr_{-}\over {r_{+}-r_{-}}}\ln{{r-r_{-}}\over 2M},
\label{eq:rst}
\end{eqnarray}
where $r_{\pm}=M\pm \sqrt{M^2-a^2}$. 

We solve the Teukolsky equation by using the Green function method. 
A solution of the Teukolsky equation which has a purely outgoing property at
infinity and a purely ingoing property at the horizon is given by 
\begin{equation}
R_{\ell m\omega}={1\over W} \left\{R^{\rm up}_{\ell m\omega}
\int^r_{r_+}dr' R^{\rm in}_{\ell m\omega}
T_{\ell m\omega}\Delta^{-2} 
+ R^{\rm in}_{\ell m\omega}\int^\infty_{r}dr' 
R^{\rm up}_{\ell m\omega} T_{\ell m\omega}\Delta^{-2}\right\}, 
\end{equation}
where the Wronskian $W$ is given by 
\begin{equation}
W=2 i \omega D_{\ell m\omega}A^{\rm in}_{\ell m\omega}. 
\end{equation}
Then, $\psi_4$ has an asymptotic property at the horizon such that 
\begin{equation}
R_{\ell m\omega}(r\rightarrow r_+)\rightarrow 
\tilde{Z}^{\rm H}_{\ell m\omega} \Delta^2 e^{-i k r^*}, 
\end{equation}
where
\begin{equation}
\tilde{Z}^{\rm H}_{\ell m\omega} ={C_{\ell m\omega}\over 
2i\omega D_{\ell m\omega}A^{\rm in}_{\ell m\omega}}
\int^{\infty}_{r_+}dr' R^{\rm up}_{\ell m\omega} T_{\ell m\omega}\Delta^{-2}. 
\label{eq:zhole}
\end{equation}

The energy flux formula is given by Teukolsky and Press\cite{ref:TP} as 
\begin{equation}
\left(dE_{\rm hole}\over dtd\Omega\right)=
\sum_{\ell m}\int d\omega {{_{2}S_{\ell m}^2}\over 2\pi} 
{{128 \omega k (k^2+4 {\tilde \epsilon}^2)(k^2+16 {\tilde \epsilon}^2)
(2 M r_+)^5}\over |C|^2}
|\tilde{Z}^{\rm H}_{\ell m\omega}|^2, \label{eq:dedtform}
\end{equation}
where ${\tilde \epsilon}=\kappa/(4r_+)$ and 
\begin{eqnarray}
|C|^2&=&(Q^2+4 a \omega m -4 a^2 \omega^2)[(Q-2)^2+36 a \omega m - 36 a^2 \omega^2]
\nonumber\\
&&+(2Q-1)(96 a^2 \omega^2-48 a \omega m)+144 \omega^2(M^2-a^2), 
\end{eqnarray}
and $Q=\lambda+2$. 

The post-Newtonian expansion of these formulas can be obtained 
by solving the Teukolsky equation 
by setting 
$\epsilon\equiv 2M\omega\sim O(v^3)$, $z=\omega r\sim O(v)$ and by assuming 
$\epsilon\ll z\ll 1$. Here, we define 
$v\equiv (M/r_0)^{1/2}$, where $r_0$ is the orbital radius of the test particle. 

The solution of the spin weighted spheroidal harmonics was given 
explicitly in a previous paper \cite{ref:TSTS} up to $O(\epsilon^2)$
which is accurate enough for the present purpose. 
In the next section, we explain the method to obtain the homogeneous solutions 
of the radial equation. 
\section{Analytic solutions of the homogeneous Teukolsky equation 
and the post-Newtonian expansion} 
New analytic representations of the homogeneous Teukolsky equation were
found by Mano, Suzuki and Takasugi. \cite{ref:MST}$^,$\cite{ref:MT} 
In this paper, we adopt variables used by Mano and Takasugi.\cite{ref:MT}
Since all the details are given in that paper, we only describe the outline 
of the method here. In this section, the value of the spin $s$ 
is assumed to be $s=-2$. 
In this method, the solutions of the Teukolsky equation are 
represented by two kinds of expansions. 
One is a series of hypergeometric functions which is defined as 
\begin{eqnarray} 
{R}_{0;s}^{\nu} 
&=&e^{-i\epsilon\kappa \tilde{x}}(\tilde{x})^{\nu+i\epsilon_+}
(\tilde{x}-1)^{-s-i\epsilon_+}
\cr  &&
\times \sum_{n=-\infty}^{\infty}
\frac{\Gamma(1-s-2i\epsilon_+)\Gamma(2n+2\nu+1)}
{\Gamma(n+\nu+1-i\tau)\Gamma(n+\nu+1-s-i\epsilon)}a_{n}^{\nu}
\cr &&  \times 
\tilde{x}^n F\left(-n-\nu-i\tau,-n-\nu-s-i\epsilon,
-2n-2\nu;\frac 1{\tilde{x}}\right),
\label{eq:hyperex}
\end{eqnarray}
where $F(a,b,c,x)$ is the hypergeometric function, and 
\begin{eqnarray} 
&&\tilde{x}={\omega\over {\epsilon\kappa}}(r-r_-),
~~~\kappa=\sqrt{1-q^2},~~~q={a\over M}, 
~~~\tau={{\epsilon-m q}\over \kappa}. 
\end{eqnarray}
The coefficients $a_n^{\nu}$ obey the three terms recurrence relation
\begin{equation}
\alpha_n^\nu a_{n+1}^\nu+\beta^\nu_n a_n^\nu+\gamma^\nu_n a_{n-1}^\nu=0, 
\label{eq:three}
\end{equation}
where
\begin{eqnarray}
\alpha_n^{\nu}&=&{i\epsilon \kappa (n+\nu+1+s+i\epsilon)
(n+\nu+1+s-i\epsilon)(n+\nu+1+i\tau)
\over{(n+\nu+1)(2n+2\nu+3)}},
\cr
\beta_n^{\nu}&=&-\lambda-s(s+1)+(n+\nu)(n+\nu+1)+\epsilon^2 
+\epsilon(\epsilon-m q)\cr
&&\hskip 3mm +{\epsilon (\epsilon-mq)(s^2+\epsilon^2) \over{(n+\nu)(n+\nu+1)}}, 
\cr
\gamma_n^{\nu}&=&-{i\epsilon \kappa
 (n+\nu-s+i\epsilon)(n+\nu-s-i\epsilon)(n+\nu-i\tau)
\over{(n+\nu)(2n+2\nu-1)}}.
\end{eqnarray}

The series converges if $\nu$ satisfies the equation
\begin{equation}
R_n(\nu)L_{n-1}(\nu)=1, \label{eq:nueq}
\end{equation}
where $R_n(\nu)$ and $L_n(\nu)$ are continued fractions defined by 
\begin{eqnarray}
R_n(\nu)&\equiv&{a_n^\nu \over a_{n-1}^\nu}=
-{\gamma_n^\nu\over {\beta_{n}^{\nu}+}}
{-\alpha_n^\nu \gamma_{n+1}^\nu\over {\beta_{n+1}^\nu+}}
{-\alpha_{n+1}^\nu \gamma_{n+2}^\nu\over {\beta_{n+2}^\nu+}}... \,,
\label{eq:cont1}
\cr
L_n(\nu)&\equiv&{a_n^\nu \over a_{n+1}^\nu}=
-{\alpha_n^\nu\over {\beta_n^\nu+}}
{-\alpha_{n-1}^\nu \gamma_{n}^\nu\over {\beta_{n-1}^\nu+}}
{-\alpha_{n-2}^\nu \gamma_{n+1}^\nu\over {\beta_{n-2}^\nu+}}...\,.
\label{eq:cont2}
\end{eqnarray}
Then the expansion equation (\ref{eq:hyperex}) converges if $r<\infty$. 

The ingoing solution $R^{\rm in}_{\ell m\omega}$ is given by
\begin{eqnarray}
R^{\rm in}_{\ell m\omega}=e^{i\epsilon\kappa}(R_0^{\nu}+R_0^{-\nu-1}). 
\end{eqnarray}

The other solution is a series of Coulomb wave functions, 
defined as \cite{ref:leaver} 
\begin{equation} 
{R}_{C;s}^{\nu}= 
\tilde{z}^{-1-s}\left(1-{\epsilon \kappa \over{\tilde{z}}}\right)^{-s-i\epsilon_+}
\sum_{n=-\infty}^{\infty}(-i)^n
\frac{(\nu+1+s-i\epsilon)_n}{(\nu+1-s+i\epsilon)_n}
a_n^{\nu}F_{n+\nu,s}(\tilde{z}), \label{eq:coulombex} 
\end{equation} 
where $(a)_n=\Gamma(n+a)/\Gamma(a)$, and
$F_{n+\nu,s}(\tilde{z})$ is the Coulomb wave function defined by 
\begin{eqnarray*} 
F_{n+\nu}(z)&=&e^{-iz}(2z)^{n+\nu} z 
{\Gamma(n+\nu+1-s+i\epsilon)\over{\Gamma(2n+2\nu+2)}} \cr
&&\times{\rm \Phi}(n+\nu+1-s+i\epsilon,2n+2\nu+2;2iz), 
\end{eqnarray*} 
where ${\rm \Phi}(a,b,z)$ is a confluent hypergeometric 
function,\cite{ref:handbook} $\tilde{z}=\omega(r-r_{-})$ 
and $\epsilon_{\pm}=(\epsilon\pm\tau)/2$. 
The expansion coefficients ${a_n^\nu}$ obey the same recurrence relation
(\ref{eq:three}). 
The expansion equation (\ref{eq:coulombex}) converges if 
$\nu$ is a solution of Eq.~(\ref{eq:nueq}). 
In that case, the series converge for $\tilde{z}>\epsilon\kappa$. 

This solution is decomposed into  
a pair of solutions, the incoming solution at infinity, $R_{+;s}^{\nu}$, 
and the outgoing solution at infinity, $R_{-;s}^{\nu}$. 
Explicitly, we have 
\begin{equation}
R_{C;s}^{\nu}=R_{+;s}^{\nu}+R_{-;s}^{\nu},
\end{equation}
where 
\begin{eqnarray}
&&R_{+;s}^{\nu}=2^{\nu}e^{-\pi \epsilon}e^{i\pi(\nu+1-s)}
\frac{\Gamma(\nu+1-s+i\epsilon)}{\Gamma(\nu+1+s-i\epsilon)}
e^{-i\tilde{z}}\tilde{z}^{\nu+i\epsilon_+}
(\tilde{z}-\epsilon\kappa)^{-s-i\epsilon_+}
\cr
&&~~~~~~\times \sum_{n=-\infty}^{\infty}i^n
a_n^{\nu}(2\tilde{z})^n{\rm \Psi}(n+\nu+1-s+i\epsilon,2n+2\nu+2;2i\tilde{z}),
\\ 
&&R_{-;s}^{\nu}= 2^{\nu}e^{-\pi \epsilon}e^{-i\pi(\nu+1+s)}
e^{i\tilde{z}}\tilde{z}^{\nu+i\epsilon_+}
(\tilde{z}-\epsilon\kappa)^{-s-i\epsilon_+}
\sum_{n=-\infty}^{\infty}i^n\cr
&&\times \frac{(\nu+1+s-i\epsilon)_n}{(\nu+1-s+i\epsilon)_n}
a_n^{\nu}(2\tilde{z})^n{\rm \Psi}(n+\nu+1+s-i\epsilon,2n+2\nu+2;-2i\tilde{z}) ,
\label{eq:Rminus}
\end{eqnarray}
where ${\rm \Psi}(a,b,z)$ is a confluent hypergeometric 
function.\cite{ref:handbook}
The upgoing solution $R^{\rm up}_{\ell m\omega}$ is then given by 
\begin{equation}
R^{\rm up}_{\ell m\omega}=R_{-;s}^{\nu} \, . 
\end{equation}

We see that the above two kinds of solutions have 
a very wide convergence region. 
We also see that the two solutions $R^{\nu}_{0;s}$ and $R^{\nu}_{C;s}$
are not linearly independent of each other. 
Then $R^{\nu}_{0;s}$ must be proportional to $R^{\nu}_{C;s}$. 
The relation between them is given by 
$$
R_{0;s}^\nu=K_\nu R^\nu_{C;s}, 
$$
where
\begin{eqnarray}
{K}_{\nu}(s)&=&
\frac{(2\epsilon \kappa )^{-\nu+s-{\tilde r}}2^{-s}i^{\tilde r} 
\Gamma(1-s-2i\epsilon_+)\Gamma({\tilde r}+2\nu+1)\Gamma({\tilde r}+2\nu+2)}
{\Gamma(\nu+1-i\tau)\Gamma(\nu+1-s-i\epsilon)\Gamma({\tilde r}+\nu+1-s+i\epsilon)} 
\cr &&
\times \frac{\Gamma(\nu+1+i\tau)\Gamma(\nu+1+s+i\epsilon)}
{\Gamma({\tilde r}+\nu+1+i\tau)\Gamma({\tilde r}+\nu+1+s+i\epsilon)}
\cr&&
\times \left ( \sum_{n={\tilde r}}^{\infty}
\frac{({\tilde r}+2\nu+1)_n}{(n-{\tilde r})!}
\frac{(\nu+1+s-i\epsilon)_n}{(\nu+1-s+i\epsilon)_n}
a_n^{\nu}(s) \right )^*
\cr &&
\times \left(\sum_{n=-\infty}^{{\tilde r}}
\frac{(-1)^n}{({\tilde r}-n)!
({\tilde r}+2\nu+2)_n}\frac{(\nu+1+s-i\epsilon)_n}{(\nu+1-s+i\epsilon)_n}
a_n^{\nu}(s)\right)^{-1}, 
\end{eqnarray}
where $\tilde r$ can be any  integer  and 
${K}_{\nu}(s)$ is independent of the choice of  ${\tilde r}$.
This relation holds in a region 
where both expansion equations (\ref{eq:hyperex}) and 
(\ref{eq:coulombex}) converge. 

Using this relation, it is possible to obtain the asymptotic amplitudes of 
$R^{\rm in}_{\ell m\omega}$ at infinity. They are given by 
\begin{eqnarray}
&&A^{\rm in}_{\ell m\omega}
=\frac{ e^{i\epsilon\kappa}}{\omega}\left[{ K}_{\nu}(s)-
 ie^{-i\pi\nu} \frac{\sin \pi(\nu-s+i\epsilon)}{\sin \pi(\nu+s-i\epsilon)}
{ K}_{-\nu-1}(s) \right]A_{+;s}^{\nu} e^{-i\epsilon\ln\epsilon}, \\
&&A^{\rm out}_{\ell m\omega}
=\frac{ e^{i\epsilon\kappa}}{\omega^{1+2s}}\left[{ K}_{\nu}(s)+ie^{i\pi\nu} 
{ K}_{-\nu-1}(s) \right]A_{-;s}^{\nu} e^{i\epsilon\ln\epsilon}, 
\end{eqnarray}
where 
\begin{equation}
A_{+;s}^{\nu}=2^{-1+s-i\epsilon}e^{i(\pi/2)(\nu+1-s)}e^{-\pi \epsilon/2}
\frac{\Gamma(\nu+1-s+i\epsilon)}{\Gamma(\nu+1+s-i\epsilon)}\sum_{n=-\infty}^{\infty} 
a_n^{\nu}, 
\end{equation}
\begin{equation}
A_{-;s}^{\nu}=2^{-1-s+i\epsilon}e^{-i(\pi/2)(\nu+1+s)}e^{-\pi \epsilon/2}
\sum_{n=-\infty}^{\infty}(-1)^n
\frac{(\nu+1+s-i\epsilon)_n}{(\nu+1-s+i\epsilon)_n}a_n^{\nu}.
\end{equation}

The asymptotic amplitude at the outer horizon  $C_{\ell m\omega}$
is given by 
\begin{equation} 
C_{\ell m\omega}=\left(\frac{\epsilon\kappa}{\omega}\right)^{2s} 
e^{i\epsilon_{+}\ln\kappa}
 \sum_{n=-\infty}^{\infty}a_{n}^{\nu}. 
\end{equation} 
The asymptotic amplitude $D_{\ell m\omega}$ of $R^{\rm up}_{\ell m\omega}$ 
at infinity is also given by 
\begin{equation} 
D_{\ell m\omega} 
=2^{-1-s+i\epsilon}e^{-\pi\epsilon/2} e^{-{\pi\over 2}i(\nu+1+s)} 
     \left[\sum_{n=-\infty}^{\infty}{(\nu+1+s-i\epsilon)_n\over 
     (\nu+1-s+i\epsilon)_n}a_n^\nu (-1)^n \right] 
     {e^{i\epsilon\ln\epsilon}\over \omega^{1+2s}}. 
\end{equation}

Next, we explain the relation of the method above to the post-Newtonian 
expansion. It is easy to see that 
if we set $a_0^\nu=1$ at $n=0$, the order of $a_n$ in $\epsilon$ 
increases with $|n|$. 
Basically, this is because $\alpha_n^\nu$ and $\gamma_n^\nu$ contain 
an overall factor $\epsilon$. 
Further, from the behavior of each term of the expansion
equation (\ref{eq:coulombex}) 
as a function of $z$, we see that 
the post-Newtonian order of the expansion equation (\ref{eq:coulombex}) 
increases. This can be checked by setting $\epsilon\sim v$ and $z\sim v$, where 
$v=(M/r_0)^{1/2}$, and $r_0$ is the orbital radius of the test particle. 
Then, the above expansion is very useful for the post-Newtonian 
expansion. Note that these increases of the post-Newtonian order 
are not monotonic when $n<0$. \cite{ref:MST} 
Therefore we must be careful to treat the expansion for $n<0$. 
This is not, however, a serious problem because 
we can obtain the desired post-Newtonian accuracy by summing up the terms 
up to the appropriately large $|n|$. 

Finally, we briefly explain the procedure of the analytic calculation. 
We first calculate the eigenvalue $\lambda$ in a power series of $\epsilon$ 
using a method due to Fackerell and Crossman.\cite{ref:fackerell} 
Next we determine $\nu$ by 
solving Eq. (\ref{eq:nueq}). This equation can also be solved in a power series
of $\epsilon$. Then we evaluate the expansion coefficients ${a_n^\nu}$. 
This can be done conveniently by 
using continued fractions Eqs. (\ref{eq:cont1}) 
and setting $a_0^\nu=1$. Once we obtain the $a_n^\nu$, 
it is straightforward to calculate asymptotic amplitudes 
$A^{\rm in}_{\ell m\omega}$, $C_{\ell m\omega}$ and $D_{\ell m\omega}$. 
It is also straightforward but tedious to 
calculate $R^{\rm up}_{\ell m\omega}$ from Eq. (\ref{eq:Rminus})
by assuming $\epsilon \ll$ $z \ll 1$. 
These results are given in Appendix B. 

\section{The black hole absorption to $O(v^8)$}
In this section, we evaluate Eq. (\ref{eq:dedtform}). 
The calculation is almost the same as those given in section III of a 
paper TSTS. 
First, we solve the geodesic equation for circular motion in the equatorial 
plane. Next, we calculate the source term $T_{\ell m\omega}$. 
The result is given by 
\begin{eqnarray}
T_{\ell m\omega}&=&\int^{\infty}_{-\infty}dt e^{i\omega t-i m \varphi(t)}
\Delta^2\bigl[(A_{n\,n\,0}+A_{{\bar m}\,n\,0}+
A_{{\bar m}\,{\bar m}\,0})\delta(r-r_0)
\cr
&+&
\left\{(A_{{\bar m}\,n\,1}+A_{{\bar m}\,{\bar m}\,1})
\delta(r-r_0)\right\}_{,r}
+\left\{A_{{\bar m}\,{\bar m}\,2}\delta(r-r_0)\right\}_{,rr}
\bigr]_{\theta=\pi/2}\,,
\label{eq:tone} 
\end{eqnarray} 
where $A_{n\,n\,0}$, etc., are given in Appendix A. 
Inserting Eq. (\ref{eq:tone}) into Eq. (\ref{eq:zhole}), we obtain 
$\tilde Z^{\rm H}_{\ell m\omega}$ as
\begin{eqnarray}
\tilde Z^{\rm H}_{\ell m\omega}&=&
{{2 \pi C_{\ell m\omega}\delta(\omega-m\Omega)}
\over {2i\omega A^{{\rm in}}_{\ell m\omega} D_{\ell m\omega}}}
\Bigl[R^{{\rm up}}_{\ell m\omega}\{A_{n\,n\,0}+A_{{\bar m}\,n\,0}
+A_{{\bar m}\,{\bar m}\,0}\}
\nonumber\\
& &-{dR^{{\rm up}}_{\ell m\omega} \over dr}\{ A_{{\bar m}\,n\,1}
+A_{{\bar m}\,{\bar m}\,1}\}
+{d^2 R^{{\rm up}}_{\ell m\omega} \over dr^2}
A_{{\bar m}\,{\bar m}\,2}\Bigr]_{r=r_0,\theta=\pi/2}
\nonumber\\
&\equiv& 
\delta(\omega-m\Omega) Z^{\rm H}_{\ell m}\,,
\label{eq:zzq}
\end{eqnarray}
where $\Omega$ is the orbital angular frequency given by 
\begin{equation}
\Omega={M^{1/2}\over r_0^{3/2}}
\left[1 - q v^3+q^2 v^6+O(v^9)\right]. 
\end{equation}
From Eqs. (\ref{eq:dedtform}) and (\ref{eq:zzq}), the time averaged 
energy absorption rate becomes
\begin{eqnarray} 
\left(dE\over dt\right)_{\rm H}&=&
\sum_{\ell m}\left[
{{128 \omega k (k^2+4 {\tilde \epsilon}^2)(k^2+16 {\tilde \epsilon}^2)
(2 M r_+)^5}\over |C|^2}
|Z^{\rm H}_{\ell m}|^2\right]_{\omega=m\Omega} \cr
&\equiv&\sum_{\ell m}\left(dE\over dt\right)_{\ell,m}. \label{eq:dedt}
\end{eqnarray}
Using a property of $_{-2}S^{a\omega}_{\ell m}(\theta)$ 
at $\theta=\pi/2$, 
it is straightforward to show that $\bar{T}_{\ell, -m, -\omega}$ 
$=(-1)^{\ell}T_{\ell,m,\omega}$ 
where $\bar{T}_{\ell, m, \omega}$ is the complex conjugate of 
$T_{\ell,m,\omega}$. 
Since the homogeneous Teukolsky equation is invariant 
under complex conjugation followed by $m\rightarrow -m$ and
$\omega\rightarrow -\omega$, 
we have $\bar{Z}_{\ell, -m, -\omega}$ 
$=(-1)^{\ell}Z_{\ell,m,\omega}$. 
Then, from Eq. (\ref{eq:dedt}), we have 
$(dE/dt)_{\ell,-m}=(dE/dt)_{\ell,m}$.

In order to express the post-Newtonian corrections to the black hole
absorption, 
we define $\eta_{\ell,m}^{\rm H}$ as 
\begin{equation}
\left( {dE \over dt} \right)_{\ell,m}
\equiv{1\over2}\left({dE\over dt}\right)_N v^5 \eta_{\ell,m}^{\rm H}\,,
\label{eq:etalmw}
\end{equation}
where $(dE/dt)_N$ is the Newtonian quadrupole luminosity at infinity,
$$
\left({dE\over dt}\right)_N={32\mu^2M^3\over5r_0^5}
={32\over5}\left({\mu\over M}\right)^2v^{10} . 
\label{eq:Enewton}
$$
In Appendix C, we give $\eta_{\ell m}$
for $m>0$. The results for $m<0$ are given by 
$\eta_{\ell,m}=\eta_{\ell,-m}$. 

The total absorption rate is given by 
\begin{eqnarray}
\left( {dE\over dt}\right)_{\rm H}=&&
\left(dE\over dt\right)_N\, v^5\, \Biggl[
-{3\over 4}\,{q}^{3}-{1\over 4}\,q
+\left (-q-{\frac {33}{16}}\,{q}^{3}\right ){v}^{2} \nonumber\\
&&
+\left ({7\over 2}\,{q}^{4}+2\,q{\it B_2}+{1\over 2}+6\,{q}^{3}{\it B_2
}+{\frac {85}{12}}\,{q}^{2}+3\,{q}^{4}\kappa+{1\over 2}\,\kappa
+{13\over 2}\,\kappa
\,{q}^{2}\right ){v}^{3} \nonumber\\
&&
+\left (-{\frac {
4651}{336}}\,{q}^{3}-{\frac {43}{7}}\,q-{\frac {17}{56}}\,{q}^{5}
\right ){v}^{4} \nonumber\\
&&
+\Bigl({\frac {569}{24}}\,{q}
^{2}+{\frac {371}{48}}\,{q}^{4}+18\,{q}^{3}{\it B_2}-{3\over 4}\,{q}^{3}{\it 
B_1}+2\,\kappa+2+{\frac {33}{4}}\,{q}^{4}\kappa \nonumber\\
&&
+6\,q{\it B_2}+{\frac {163}{8}}\,\kappa\,{q}^{2}+q{\it B_1}\Bigr){v}^{5} 
\nonumber\\
&&
+\Bigl(-{
\frac {2718629}{44100}}\,q-4\,{\it B_2}+{\frac {428}{105}}\,\gamma\,q
+{2\over 3}\,{\pi }^{2}q+{\frac {428}{105}}\,q\ln 2-4\,q{\it C_2} \nonumber\\
&&
-12\,{q}^{3}{\it C_2}-36\,{q}^{4}{\it B_2}-56\,{q}^{2}{\it B_2}
+{\frac {428}{35}}\,{q}
^{3}\gamma+{\frac {428}{35}}\,{q}^{3}\ln 2+2\,{q}^{3}{\pi }^{2} \nonumber\\
&&
+{\frac {428}{105}}\,q\ln\kappa+{\frac {428}{105}}\,q{\it A_2}+{\frac 
{428}{35}}\,{q}^{3}\ln\kappa+6\,{\frac {{q}^{7}}{\kappa}}+{\frac {
428}{35}}\,{q}^{3}{\it A_2}-8\,q{{\it B_2}}^{2} \nonumber\\
&&
-24\,{q}^{3}{{\it B_2}}^{2}+{\frac {856}{105}}\,q\ln v
+{\frac {856}{35}}\,{q}^{3}\ln v-4\,{
\frac {{\it B_2}}{\kappa}}-32\,{\frac {{q}^{3}}{\kappa}}-31\,{\frac {q}
{\kappa}} \nonumber\\
&&
+57\,{\frac {{q}^{5}}{\kappa}}+{q}^{5}\kappa+{q}^{4}{\it B_1}-
{2\over 3}\,\kappa\,q-{7\over 6}\,\kappa\,{q}^{3}
-{4\over 3}\,{q}^{2}{\it B_1} \nonumber\\
&&
-48\,{\frac {{q}^{2}{\it B_2}}{\kappa}}
+28\,{\frac {{q}^{4}{\it B_2}}{\kappa}}-24\,{
\frac {{q}^{3}{\it C_2}}{\kappa}}-8\,{\frac {q{\it C_2}}{\kappa}}+24\,{
\frac {{q}^{6}{\it B_2}}{\kappa}} \nonumber\\
&&
-{\frac {2400247}{19600}}\,{q}^{3}+{
\frac {299}{16}}\,{q}^{5}\Bigr){v}^{6} \nonumber\\
&&
+\Bigl({\frac {225}{28}}\,{q}^{5}{\it B_3}-{\frac {41}{28}}\,\kappa\,
{q}^{6}+{\frac {86}{7}}+{\frac {8741}{56}}\,{q}^{2}+{\frac {3485}{42}}
\,{q}^{4}+{\frac {167}{112}}\,{q}^{6} \nonumber\\
&&
+{\frac {86}{7}}\,\kappa+{\frac {
45}{56}}\,q{\it B_3}-{\frac {9}{28}}\,{q}^{5}{\it B_1}+{\frac {899}{168}
}\,q{\it B_1}-{\frac {803}{224}}\,{q}^{3}{\it B_1}+{\frac {1665}{224}}\,
{q}^{3}{\it B_3} \nonumber\\
&&
+{\frac {2372}{21}}\,{q}^{3}{\it B_2}-{\frac {16}{7}}\,{
q}^{5}{\it B_2}+{\frac {719}{12}}\,{q}^{4}\kappa+{\frac {22201}{168}}\,
\kappa\,{q}^{2}+{\frac {796}{21}}\,q{\it B_2}\Bigr){v}^{7} 
\nonumber\\
&&
+\Bigl(-{\frac {20542807}{88200}}\,q-12\,{\it B_2}-2\,{\it B_1}+{\frac {
1061}{35}}\,\gamma\,q+{\frac {13}{6}}\,{\pi }^{2}q+{\frac {995}{21}}\,
q\ln 2\nonumber\\
&&
-12\,q{\it C_2}-36\,{q}^{3}{\it C_2}-{\frac {308}{3}}\,{q}^{4}{
\it B_2}-{\frac {1496}{9}}\,{q}^{2}{\it B_2}+{\frac {12197}{140}}\,{q}^{
3}\gamma+{\frac {3873}{28}}\,{q}^{3}\ln 2 \nonumber\\
&&
+{\frac {47}{8}}\,{q}^{3}{\pi }^{2}+{\frac {1391}{105}}\,q\ln\kappa
+{\frac {428}{35}}\,q{\it A_2}+{\frac {5029}{140}}\,{q}^{3}\ln\kappa
+{\frac {37}{6}}\,{\frac {{q}^{7}}{\kappa}}
+{\frac {1284}{35}}\,{q}^{3}{\it A_2} \nonumber\\
&&
-24\,q{{\it B_2}}^{2}-72\,{q}^{3}{{\it B_2}}^{2}
+{\frac {4574}{105}}\,q\ln v+{\frac {8613}{70}}\,{q}^{3}\ln v
-12\,{\frac {{\it B_2}}{\kappa}}-{\frac {341}{4}}\,{\frac {{q}^{3}}{\kappa}} 
\nonumber\\
&&
-{\frac {637}{6}}\,{\frac {q}{\kappa}}+
{\frac {741}{4}}\,{\frac {{q}^{5}}{\kappa}}+{\frac {73}{6}}\,{q}^{4}{
\it B_1}-q{\it C_1}-{\frac {283}{18}}\,{q}^{2}{\it B_1} \nonumber\\
&&
+{3\over 2}\,{q}^{3}{{\it B_1}}^{2}+{\frac {107}{105}}\,q{\it A_1}
-{\frac {107}{140}}\,{q}^{3}
{\it A_1}-3\,{\frac {{q}^{6}{\it B_1}}{\kappa}}+{13\over 2}\,{\frac {{q}^{4}{
\it B_1}}{\kappa}}-{3\over 2}\,{\frac {{q}^{2}{\it B_1}}{\kappa}} \nonumber\\
&&
-2\,{\frac {q{\it C_1}}{\kappa}}+{3\over 2}\,{\frac {{q}^{3}{\it C_1}}{\kappa}}
-2\,q{{\it B_1}
}^{2}+{3\over 4}\,{q}^{3}{\it C_1}-2\,{\frac {{\it B_1}}{\kappa}}-144\,{\frac {
{q}^{2}{\it B_2}}{\kappa}} \nonumber\\
&&
+84\,{\frac {{q}^{4}{\it B_2}}{\kappa}}-72\,{
\frac {{q}^{3}{\it C_2}}{\kappa}}-24\,{\frac {q{\it C_2}}{\kappa}}+72\,{
\frac {{q}^{6}{\it B_2}}{\kappa}}-{\frac {2945984497}{6350400}}\,{q}^{3}
\nonumber\\
&&
+{\frac {1385}{24}}\,{q}^{5}+{\frac {25}{252}}\,{q}^{7}\Bigr){v}^{8}
\Biggr], \label{eq:dEdt} 
\end{eqnarray}
where 
\begin{eqnarray}
A_n&=&{1\over 2}\left[\psi^{(0)}\left(3+{n i q\over \sqrt{1-q^2}}\right)
+\psi^{(0)}\left(3-{n i q\over \sqrt{1-q^2}}\right)
\right], \cr
B_n&=&{1\over 2i}\left[\psi^{(0)}\left(3+{n i q\over \sqrt{1-q^2}}\right)
-\psi^{(0)}\left(3-{n i q\over \sqrt{1-q^2}}\right)
\right], \cr
C_n&=&{1\over 2}\left[\psi^{(1)}\left(3+{n i q\over \sqrt{1-q^2}}\right)
+\psi^{(1)}\left(3-{n i q\over \sqrt{1-q^2}}\right)
\right], 
\end{eqnarray}
and $\psi^{(n)}(z)$ is the polygamma function. 

We see that 
the absorption effect starts at $O(v^5)$ beyond the quadrapole formula 
in the case $q\neq 0$. 
If we set $q=0$, the above formula reduces to 
\begin{equation}
\left({dE\over dt}\right)_{\rm H}=
\left(dE\over dt\right)_N \left(v^8+O(v^{10})\right), 
\end{equation}
which was derived by Poisson and Sasaki.\cite{ref:PoSa}
Then, the black hole absorption is more important in the case 
$q\neq 0$. 
We also notice that Eq. (\ref{eq:dEdt}) can be negative when $q>0$. 
This is an example of super radiance phenomenon. 

It is not clear that Eq. (\ref{eq:dEdt}) has a finite limit as 
$|q|\rightarrow 1$. However by using the formulas 
\begin{equation}
\lim_{q\rightarrow \pm 1}
\psi^{(0)}\left(3+{n i q\over \sqrt{1-q^2}}\right)
=\ln{n}-\ln\kappa+i{q\over |q|}{\pi\over 2}, 
\end{equation}
\begin{equation}
\lim_{q\rightarrow \pm 1}
\psi^{(k)}\left(3+{n i q\over \sqrt{1-q^2}}\right)=0, 
~~~(k\neq 0)
\end{equation}
we can obtain the limit of $(dE/dt)_{\rm H}$ as 
\begin{eqnarray}
\lim_{q\rightarrow \pm 1}\left(dE\over dt\right)_{\rm H}
=&&\left(dE\over dt\right)_N v^5 \Biggl[
-{q\over |q|}
-{\frac {49}{16}}\,{q\over |q|}\,{v}^{2}
+\left (4\,\pi +{\frac {133}{12}}\right ){v}^{3}
\nonumber \\
&&
-{\frac {6817}{336}}\,{q\over |q|}\,{v}^{4}
+\left ({\frac {535}{16}}+{\frac {97}{8}}\,\pi \right ){v}^{5}
\nonumber \\
&&
+\left ({\frac {3424}{105}}\,{q\over |q|}\,\ln (2)
+{\frac {1712}{105}}\,\gamma\,{q\over |q|}
+{\frac {3424}{105}}\,{q\over |q|}\,\ln (v)
\right.
\nonumber \\
&&
\left.
-{\frac {3647533}{22050}}\,{q\over |q|}
-{\frac {289}{6}}\,{q\over |q|}\,\pi 
-16/3\,{\pi }^{2}{q\over |q|}\right ){v}^{6}
\nonumber \\
&&
+\left ({\frac {84955}{336}}+{\frac {55873}{672}}\,\pi \right ){v}^{7}
\nonumber \\
&&
+\left ({\frac {14077}{60}}\,{q\over |q|}\,\ln (2)+{\frac {16441}{140}}\,
\gamma\,{q\over |q|}+{\frac {34987}{210}}\,{q\over |q|}\,\ln (v)-{\frac {
193}{12}}\,{\pi }^{2}{q\over |q|}
\right.
\nonumber \\
&& 
\left.
-{\frac {4057965601}{6350400}}\,{q\over |q|}
-{\frac {1289}{9}}\,{q\over |q|}\,\pi \right ){v}^{8}\Biggr]. 
\end{eqnarray}
In Appendix D, we give $(dE/dt)_{\rm H}$ written in terms of 
$x\equiv(M\Omega)^{1/3}$.

\section{The orbital evolution of coalescing compact binaries}

Using the above results, we estimate the effect of black hole absorption 
on the orbital evolution of the coalescing compact binaries. 
We ignore the finite mass effect in the post-Newtonian formulas and 
interpret $M$ as the total mass and $\mu$ as the reduced mass 
of the system. 

The total cycle $N$ of gravitational waves from an inspiraling 
binary is given by 
\begin{equation} 
N=\int^{v_i}_{v_f}dv {\Omega\over \pi} {dE/dv\over dE/dt}, 
\end{equation} 
where $v_i=(M/r_i)^{1/2}$, $v_f=(M/r_f)^{1/2}$, 
and $r_i$ and $r_f$ are the initial and final orbital separation of the 
binary. The energy loss rate is given by 
$dE/dt=$ $(dE/dt)_{\rm H}+$ $(dE/dt)_\infty$, where $(dE/dt)_\infty$ is the 
luminosity at infinity.

We set $r_f=6 M$ and define $r_i$ to be the radius at which the frequency of 
the wave is 10Hz. 
We evaluate $\Delta N$ which is the difference of the total cycle $N$ caused by 
the formulas for $dE/dt$ with and without the black hole absorption effect. 
The value of $\Delta N$ is calculated explicitly by 
\begin{equation}
\Delta N=\left|\int^{v_i}_{v_f}dv {\Omega\over \pi} 
{dE/dv\over {(dE/dt)_{\rm H}+(dE/dt)_\infty}}
-\int^{v_i}_{v_f}dv {\Omega\over \pi} {dE/dv\over (dE/dt)_\infty}\,\right|. 
\end{equation}

The results for the typical NS-BH systems are given in Table I. 
We find that the effect of black hole absorption is very small 
when $q=0$. We also find that the black 
hole absorption is more important
in cases for which $q>0$ than in cases for which $q<0$.
This is because, when $q<0$ 
the total cycle is much smaller than the case $q>0$. 
We also notice that 
the black hole absorption is more important when the mass of the
black hole is large. The reason is as follows. \cite{ref:TTS} 

\begin{table}
\caption{The difference of the total cycle $N$ caused by 
the formulas for $dE/dt$ with and without the black hole absorption effect 
for typical neutron star(NS)-black hole(BH) binaries with mass 
$(M_{\rm NS},M_{\rm BH})$. }
\begin{center}
\begin{tabular}{cccc} \hline\hline
\large q &\large (1.4$M_\odot$,10$M_\odot$)&\large(1.4$M_\odot$,40$M_\odot$)& 
\large(1.4$M_\odot$,70$M_\odot$)\\  \hline
\large 0  &\large $<10^{-2}$ &\large $0.02$ &\large $0.1$  \\
\large 0.9    &\large $0.5$      &\large $1.1$    &\large $1.5$  \\
\large $-0.9$ &\large $0.3$      &\large $0.5$ &\large $0.4$ 
\\ \hline
\end{tabular}
\end{center}
\end{table}

When the black hole mass becomes larger, 
the initial non-dimensional orbital radius $r_i/M$ becomes smaller 
and thus $v_i$ becomes larger. When $q>0$, this effect is dominant. 
Hence, although the total cycle $N$ becomes small when 
the black hole mass is large, 
the effect of the black hole absorption becomes larger. 

\section{Summary}

We have calculated expressions for the gravitational waves 
which are absorbed by a rotating black hole and 
obtained the energy absorption rate by the black hole 
to $O(v^{13})$ beyond the Newtonian 
quadrapole luminosity at infinity. We adopted a method developed by 
Mano et al.\cite{ref:MST} to 
calculate the post-Newtonian expansion of the Teukolsky equation analytically. 
We find that black hole absorption occurs at 
$O(v^5)$ beyond the quadrapole formula. This order is $O(v^3)$ lower than in the
non-rotating case. 
The energy absorption rate obtained here displays a property of super radiance. 

Using the formulas obtained above, we have estimated the effect of the 
black hole absorption on the orbital evolution of coalescing 
neutron star-black hole binaries. 
We calculated its effect on the cycle of waves from the binaries
in the laser interferometer's band. 
We found that the effect of the black hole absorption is 
more important in the case when $q>0$. 
(This is the case when a particle moves 
in the direction of the black hole rotation.) 
We also found that black hole absorption is important 
in the case when the mass of black hole is large. 

Gal'tsov\cite{ref:galtsov} gave formulas for black hole 
absorption from the radiation reaction force on the particle.
Gal'tsov's formula is expressed as 
\begin{eqnarray}
{dE \over dt}=\biggl( {dE \over dt} \biggr)_N {v^5 \over 2}
\left\{v^3 \Bigl(1+\sqrt{1-q^2}\Bigr)
-{q \over 2} \right\}\left(1+3q^2\right). 
\end{eqnarray}
This formula agrees with our formula Eq. (\ref{eq:dEdt}) 
to lowest order. 

In this paper, we have calculated the post-Newtonian formulas only to 
$O(v^8)$ beyond the lowest order. 
However, it is straightforward to calculate higher orders with 
our method. 

The analysis in this paper has been restricted to the case
in which a test particle 
moves in a circular orbit on the equatorial plane. 
However, it has been suggested\cite{ref:TSTS} that 
the inclination of the orbital plane from the equatorial plane will  
significantly affect the orbital phase evolution. 
Hence, the effect of the orbital inclination should be taken into account
in future works. 

\begin{center}
\Large{\bf Acknowledgements}
\end{center}

The authors thank M. Sasaki for useful discussion and 
for T. Nakamura, M. Shibata and 
T. Tanaka for stimulating conversations. 
H. Tagoshi was supported by Research Fellowships of 
the Japan Society for the Promotion of the 
Science for Young Scientists. 
\appendix
\section{The Functions in $Z^{\rm H}_{\ell m\omega}$}
In this appendix, we give functions which appear in $Z^{\rm H}_{\ell m\omega}$. 

The functions $A$ in Eq. (\ref{eq:tone}) are given by 
\begin{eqnarray*}
A_{n\,n\,0}&=&{-2 \over \sqrt{2\pi}\Delta^2}
C_{n\,n}\rho^{-2}{\bar \rho}^{-1}
L_1^+\{\rho^{-4}L_2^+(\rho^3 S)\},\\
A_{{\bar m}\,n\,0}&=&{2 \over \sqrt{\pi}\Delta} 
C_{{\bar m}\,n}\rho^{-3}
\Bigl[\left(L_2^+S\right)
\Bigl({iK \over \Delta}+\rho+{\bar \rho}\Bigr)
\\
& &\qquad-a\sin\theta S {K \over \Delta}({\bar \rho}-\rho)\Bigr],\\
A_{{\bar m}\,{\bar m}\,0}
&=&-{1 \over \sqrt{2\pi}}\rho^{-3}{\bar \rho}
C_{{\bar m}\,{\bar m}}S\Bigl[
-i\Bigl({K \over \Delta}\Bigr)_{,r}-{K^2 \over \Delta^2}+
2i\rho {K \over \Delta}\Bigr],\\
A_{{\bar m}\,n\,1}&=&{2\over \sqrt{\pi}\Delta }
\rho^{-3}C_{{\bar m}\,n}
[L_2^+S+ia\sin\theta({\bar \rho}-\rho)S],\\
A_{{\bar m}\,{\bar m}\,1}
&=&-{2 \over \sqrt{2\pi}}
\rho^{-3}{\bar \rho}
C_{{\bar m}\,{\bar m}}S\Bigl(i{K \over \Delta}+\rho\Bigr),\\
A_{{\bar m}\,{\bar m}\,2}
&=&-{1\over \sqrt{2\pi}}\rho^{-3}{\bar \rho}
C_{{\bar m}\,{\bar m}}S,\\
\end{eqnarray*}
where $S$ denotes $_{-2}S_{\ell m}^{a\omega}$.

The functions $C$ in the above formulas are given by 
\begin{eqnarray}
C_{n\,n}&=&{\mu \over 4\Sigma^3 \dot t}\left[E(r^2+a^2)-al_z
\right]^2,\nonumber\\
C_{{\bar m}\,n}&=&
-{\mu \rho \over 2\sqrt{2}\Sigma^2 \dot t}\left[E(r^2+a^2)-al_z
\right]
\left[i\sin\theta\Bigl(aE-{l_z \over \sin^2\theta}\Bigr)\right], 
\label{eq:cnn}\\
C_{{\bar m}\,{\bar m}}&=&
{\mu \rho^2 \over 2\Sigma \dot t }
\left[i\sin\theta 
\Bigl(aE-{l_z \over \sin^2\theta}\Bigr)\right]^2,\nonumber\\
\nonumber
\end{eqnarray}
and $\dot{t}$ is given by the geodesic equations of the particle as 
\begin{equation}
r_0^2\dot{t}=
-\Bigl(aE-{l_z}\Bigr)a
+{r^2+a^2 \over \Delta}\Bigl(E(r^2+a^2)-al_z\Bigr). 
\end{equation}
The specific energy $E$ and the angular momentum $l_z$ are given by 
\begin{eqnarray}
E&=&{{1-2v^2+q v^3}\over (1-3v^2+2qv^3)^{1/2}}\,,\cr
l_z&=&{{r_0v(1-2qv^3+q^2 v^4)}\over 
(1-3v^2+2qv^3)^{1/2}}\,. 
\end{eqnarray}

\newpage
\section{The Functions in {\it \S} 3}

\noindent
(a) $\ell=2$

\begin{eqnarray*}
A^{\rm in}_{\ell m\omega}&=&
{1\over \omega}{1\over \kappa^4 \epsilon^4}
e^{{1\over 2}\,\pi i \,\left(\nu+3\right)} e^{i \epsilon \kappa} 
e^{-i\epsilon\ln\epsilon}\\
&&
\Biggl\{ {\frac {15}{4}}
+\Biggl(-{15\over 2}\,{i}\gamma-{\frac {25}{12
}}\,mq-{\frac {15}{8}}\,\pi -{\frac {15}{4}}\,{i}
{\psi^{(0)}}\bigl(3+{\frac {{i}mq}{\kappa}}\bigr)
-{\frac {15}{4}}\,{i}\ln (2)+{\frac {125}{16}}\,{i}\Biggr){\epsilon} \\
&&
+\Biggl({\frac {1089}{56}}\,\gamma+{\frac {725}{2352}}\,{\kappa}^{2}
+{\frac {1089}{112}}\,\ln (2)-{15\over 2}\,{\gamma}^{2}
+{\frac {12625}{21168}}\,{m}^{2}{q}^{2}
+{\frac {35}{32}}\,{\pi }^{2} \\
&&
-{\frac {535}{144}}\,{i}mq
+{\frac {15}{4}}\,{i}\gamma\,\pi -{\frac {125}{32}}\,{i}\pi 
+{\frac {25}{6}}\,{i}\gamma\,mq+{\frac {25}{24}}\,m\pi \,q
-{15\over 2}\,\gamma\,\ln (2) \\
&&
-{15\over 2}\,\gamma\,\psi^{(0)}\bigl(3+{\frac {{i}mq}{\kappa}}\bigr)
-{\frac {15}{4}}\,\ln (2){\psi^{(0)}}\bigl(3+{\frac {{i}mq}{\kappa}}\bigr)
+{\frac {107}{56}}\,\ln ({\epsilon})+{\frac {107}{56}}\,\ln (\kappa) \\
&&
-{\frac {20573}{960}}-{\frac {15}{8}}\,\left (\ln (2)\right )^{2}
+{\frac {1089}{112}}\,{\psi^{(0)}}\bigl(3+{\frac {{i}mq}{\kappa}}\bigr)
-{\frac {15}{8}}\,\left ({\psi^{(0)}}
\bigl(3+{\frac {{i}mq}{\kappa}}\bigr)\right )^{2} \\
&&
-{\frac {15}{8}}\,\psi^{(1)}\bigl(3+{\frac {{i}mq}{\kappa}}\bigr)
+{\frac {25}{12}}\,{i}mq\ln (2) 
+{\frac {25}{12}}\,{i}mq{\psi^{(0)}}\bigl(3+{\frac {{i}mq}{\kappa}}\bigr) \\
&&
+{\frac {15}{8}}\,{i}\pi \,\ln (2)
+{\frac {15}{8}}\,{i}\pi \,{\psi^{(0)}}\bigl(3+{\frac {{i}mq}{\kappa}}\bigr)
-{\frac {15}{4}}\,\psi^{(1)}\bigl(3+{\frac {{i}mq}{\kappa}}\bigr){\kappa}^{-1}
\Biggr){{\epsilon}}^{2} \Biggr\}\,,
\end{eqnarray*}

\begin{eqnarray*}
C_{\ell m\omega}&=&\left(\omega\over \epsilon\kappa\right)^{4}
e^{i\epsilon_{+}\ln\kappa}
\Biggl\{
1+\left ({5\over 6}\,{i}\kappa-{\frac {5}{18}}\,mq\right )\epsilon \\
&&
+\left ({\frac {325}{7938}}\,{m}^{2}{q}^{2}+{\frac {5}{18}}-{\frac {
15}{49}}\,{\kappa}^{2}-{\frac {85}{378}}\,{i}\kappa\,mq\right )
{\epsilon}^{2} \Biggr\}\,,
\end{eqnarray*}

\begin{eqnarray*}
D_{\ell m\omega}&=&\omega^3 2^{i\epsilon} e^{-{\pi\over 2}i(\nu-1)}
e^{i\epsilon\ln\epsilon}
\Biggl\{
2+\left (-{1\over 3}\,{i}\kappa-\pi +{1\over 9}\,mq\right )\epsilon \\
&&
+\Biggl({1\over 4}\,{\pi }^{2}-{\frac {1}{189}}\,{i}\kappa\,mq
+{\frac {2}{49}}\,{q}^{2}-{1\over 3}\,{i}mq
-{\frac {11}{3969}}\,{m}^{2}{q}^{2} \\
&&
-{1\over 18}\,m\pi \,q+{1\over 6}\,{i}\kappa\,\pi -{\frac {67}{441}}
\Biggr){\epsilon}^{2} \Biggr\}\,,
\end{eqnarray*}

\begin{eqnarray*}
R^{\rm up}_{\ell m\omega}&=&
-3\,{\frac {{i}}{z}}-3+{3\over 2}\,{i}z+{1\over 2}\,{z}^{2}
-{1\over 8}\,{i}{z}^{3}
+{\frac {31}{40}}\,{z}^{4}+{\frac {43}{80}}\,{i}{z}^{5}
-{\frac {117}{560}}\,{z}^{6}-{\frac {769}{13440}}\,{i}{z}^{7}
\\
&&
+\epsilon\,\Biggl(
\Bigl({-{3\over 2}\,{i}+mq}\Bigr){1\over {z}^{2}} 
+\left({-3\,\gamma+\kappa+3\,{i}\pi -{1\over 6}\,{i}mq}\right){1\over {z}} 
-{9\over 4}\,{i}+3\,{i}\gamma
\\
&&
-{i}\kappa+3\,\pi +{1\over 3}\,mq
+\Bigl (-{5\over 2}+{3\over 2}\,\gamma-{1\over 2}\,\kappa-{3\over 2}\,{i}\pi 
-{1\over 4}\,{i}mq\Bigr )z 
\\
&&
+\Bigl ({\frac {85}{48}}\,{i}-{1\over 2}\,{i}\gamma
+{1\over 6}\,{i}\kappa-{1\over 2}\,\pi -{\frac {7}{72}}\,mq\Bigr ){z}^{2} 
\\
&&
+\Bigl (-{\frac {13}{60}}-{1\over 8}\,\gamma+{1\over 24}\,\kappa+{1\over 8}\,{i}\pi 
-{\frac {269}{720}}\,{i}mq\Bigr ){z}^{3}
\\
&&
+\Bigl (-{\frac {1559}{2400}}\,{i}+{1\over 40}\,{i}\gamma
+{\frac {31}{120}}\,{i}\kappa-{3\over 8}\,\pi +{\frac {49}{180}}\,mq
+{4\over 5}\,{i}\ln (2)+{4\over 5}\,{i}\ln (z)\Bigr ){z}^{4}\Biggr) \\
&&
+{\epsilon}^{2}\Biggl(
\Bigl(-{3\over 4}\,{i}-{\frac {3}{28}}\,{i}{\kappa}^{2}+mq
+{\frac {11}{56}}\,{i}{m}^{2}{q}^{2}\Bigr){z}^{-3} 
+\Bigl(-1-{3\over 2}\,\gamma+{1\over 2}\,\kappa+{1\over 14}\,{\kappa}^{2}
\\
&&
+{3\over 2}\,{i}\pi 
-{\frac {7}{12}}\,{i}mq-{i}\gamma\,mq+{1\over 3}\,{i}\kappa\,mq 
-m\pi \,q-{\frac {31}{504}}\,{m}^{2}{q}^{2}\Bigr){z}^{-2} 
\\
&&
+\Bigl({\frac {183}{28}}\,{i}-{\frac {107}{70}}\,{i}\gamma 
+{3\over 2}\,{i}{\gamma}^{2}-{i}\gamma\,\kappa
-{\frac {4}{49}}\,{i}{\kappa}^{2}+3\,\gamma\,\pi 
-\kappa\,\pi -{7\over 4}\,{i}{\pi }^{2}
-{1\over 4}\,mq
\\
&&
-{1\over 6}\,\gamma\,mq
+{\frac {19}{252}}\,\kappa\,mq+{1\over 6}\,{i}m\pi \,q 
+{\frac {781}{21168}}\,{i}{m}^{2}{q}^{2}-{\frac {107}{70}}\,{i}\ln (2)
\\
&&
-{\frac {107}{70}}\,{i}\ln (z)\Bigr){z}^{-1}
+{\frac {1791}{280}}-{\frac {529}{140}}\,\gamma
+{3\over 2}\,{\gamma}^{2}+{3\over 4}\,\kappa-\gamma\,\kappa
-{\frac {11}{392}}\,{\kappa}^{2}+{9\over 4}\,{i}\pi 
\\ &&
-3\,{i}\gamma\,\pi
+{i}\kappa\,\pi -{7\over 4}\,{\pi }^{2}
+{\frac {13}{24}}\,{i}mq-{1\over 3}\,{i}\gamma\,mq
+{\frac {23}{252}}\,{i}\kappa\,mq-{1\over 3}\,m\pi \,q 
\\ &&
-{\frac {563}{21168}}\,{m}^{2}{q}^{2}
-{\frac {107}{70}}\,\ln (2)
-{\frac {107}{70}}\,\ln (z)
+\Bigl (-{\frac {13751}{3360}}\,{i} 
\\ &&
+{\frac {457}{140}}\,{i}\gamma-{3\over 4}\,{i}{\gamma}^{2}
-{5\over 6}\,{i}\kappa+{1\over 2}\,{i}\gamma\,\kappa 
+{\frac {17}{4704}}\,{i}{\kappa}^{2}+{5\over 2}\,\pi 
\\ &&
-{3\over 2}\,\gamma\,\pi +{1\over 2}\,\kappa\,\pi 
+{\frac {7}{8}}\,{i}{\pi }^{2}+{2\over 3}\,mq 
-{1\over 4}\,\gamma\,mq
\\ &&
+{\frac {37}{504}}\,\kappa\,mq
+{1\over 4}\,{i}m\pi \,q+{\frac {751}{28224}}\,{i}{m}^{2}{q}^{2} 
+{\frac {107}{140}}\,{i}\ln (2)+{\frac {107}{140}}\,{i}\ln (z)\Bigr )z 
\Biggr) 
\\ &&
+{\epsilon}^{3}\Biggl(\Bigl(-{3\over 8}\,{i}
-{\frac {9}{56}}\,{i}{\kappa}^{2}+{3\over 4}\,mq
+{\frac {15}{112}}\,{\kappa}^{2}mq
+{\frac {33}{112}}\,{i}{m}^{2}{q}^{2} \\
&&
-{\frac {3}{112}}\,{m}^{3}{q}^{3}\Bigr){z}^{-4}
+\Bigl(-1-{3\over 4}\,\gamma+{1\over 4}\,\kappa-{1\over 7}\,{\kappa}^{2}
-{\frac {3}{28}}\,\gamma\,{\kappa}^{2}+{1\over 28}\,{\kappa}^{3} \\
&&
+{3\over 4}\,{i}\pi +{\frac {3}{28}}\,{i}{\kappa}^{2}\pi 
-{\frac {157}{168}}\,{i}mq-{i}\gamma\,mq+{1\over 3}\,{i}\kappa\,mq
-{\frac {5}{84}}\,{i}{\kappa}^{2}mq-m\pi \,q \\
&&
+{\frac {41}{504}}\,{m}^{2}{q}^{2}+{\frac {11}{56}}\,\gamma\,{m}^{2}{q}^{2}
-{\frac {11}{168}}\,\kappa\,{m}^{2}{q}^{2}
-{\frac {11}{56}}\,{i}{m}^{2}\pi \,{q}^{2} \\
&&
-{\frac {23}{1008}}\,{i}{m}^{3}{q}^{3}\Bigr){z}^{-3}
+\Bigl(-{1\over 3}\,{i}\kappa\,m\pi \,q
+{\frac {31}{504}}\,{i}\gamma\,{m}^{2}{q}^{2} \\
&&
-{\frac {1}{72}}\,{i}\kappa\,{m}^{2}{q}^{2}+{i}\gamma\,m\pi \,q
-{1\over 2}\,{\gamma}^{2}mq+{\frac {7}{12}}\,m{\pi }^{2}q
+{\frac {31}{504}}\,{m}^{2}\pi \,{q}^{2}+{\frac {107}{210}}\,mq\ln (2) \\
&&
+{\frac {107}{210}}\,mq\ln (z)+{\frac {2335}{42336}}\,{i}{m}^{2}{q}^{2}
-{1\over 2}\,{i}\gamma\,\kappa+{1\over 3}\,\gamma\,\kappa\,mq
+{\frac {7}{12}}\,{i}m\pi \,q-{\frac {25}{168}}\,{i}\kappa\,{q}^{2} \\
&&
-{1\over 14}\,{i}\gamma\,{\kappa}^{2}-{\frac {239}{14112}}\,{\kappa}^{2}mq+\pi 
+{\frac {1269}{280}}\,{i}+{3\over 2}\,\gamma\,\pi -{1\over 2}\,\kappa\,\pi 
-{1\over 14}\,{\kappa}^{2}\pi +{\frac {25}{504}}\,m{q}^{3} \\
&&
+{\frac {69}{196}}\,{i}{\kappa}^{2}+{\frac {33}{140}}\,{i}\gamma
-{\frac {31}{168}}\,{i}\kappa-{\frac {107}{140}}\,{i}\ln (2)
-{\frac {107}{140}}\,{i}\ln (z)+{3\over 4}\,{i}{\gamma}^{2} \\
&&
-{\frac {7}{8}}\,{i}{\pi }^{2}-{1\over 8}\,{i}{\kappa}^{3}
-{\frac {31}{420}}\,\gamma\,mq+{\frac {103}{504}}\,\kappa\,mq 
-{\frac {421}{210}}\,mq-{\frac {239}{127008}}\,{m}^{3}{q}^{3}\Bigr){z}^{-2}
\Biggr) \\
&&
+{\epsilon}^{4}\Bigl(\bigl(-{3\over 16}\,{i}-{\frac {9}{56}}\,{i}{\kappa}^{2}
-{\frac {1}{112}}\,{i}{\kappa}^{4}+{1\over 2}\,mq
+{\frac {15}{56}}\,{\kappa}^{2}mq+{\frac {33}{112}}\,{i}{m}^{2}{q}^{2} \\
&&
+{\frac {11}{252}}\,{i}{\kappa}^{2}{m}^{2}{q}^{2}
-{\frac {3}{56}}\,{m}^{3}{q}^{3}
-{\frac {23}{8064}}\,{i}{m}^{4}{q}^{4}\Bigr ){z}^{-5}\Bigr) 
\end{eqnarray*}

\noindent
(b) $\ell=3$

\begin{eqnarray*}
A^{\rm in}_{\ell m\omega}&=&
{1\over \omega}{1\over \kappa^5 \epsilon^5}
e^{{1\over 2}\,\pi i \,\left(\nu+3\right)} e^{i \epsilon \kappa} 
e^{-i\epsilon\ln\epsilon}
\Biggl\{
{\frac {945}{2}}\,{\frac {\kappa}{3\,\kappa+{i}mq}} \\
&&
-{\frac {63}{8}}\,{\frac {\kappa}{(3\,\kappa+{i}mq)^2}} 
\Biggl(90\,\kappa\,\pi +30\,{i}mq\pi 
+180\,{i}\kappa\,\ln (2)-60\,mq\ln (2)\\
&&
+45\,\kappa\,mq 
+15\,{i}{m}^{2}{q}^{2}+360\,{i}\kappa\,\gamma
-120\,\gamma\,mq-591\,{i}\kappa
+197\,mq \\
&&
+180\,{i}{\psi^{(0)}}\bigl({\frac {3\,\kappa+{i}mq}{\kappa}}\bigr)\kappa 
-60\,{\psi^{(0)}}\bigl({\frac {3\,\kappa+{i}mq}{\kappa}}\bigr)mq
-60\,{i}\Biggr)\epsilon\Biggl\} \,,
\end{eqnarray*}

\begin{eqnarray*}
C_{\ell m\omega}&=&
\left(\omega\over \epsilon\kappa\right)^{4}
e^{i\epsilon_{+}\ln\kappa}
\left (1+\left (-{\frac {11}{72}}\,mq+{2\over 3}\, i 
\kappa\right )\epsilon\right ) \,,
\end{eqnarray*}

\begin{eqnarray*}
D_{\ell m\omega}=
{\omega}^{3}2^{i\epsilon} e^{-{\pi\over 2}i(\nu-1)}
e^{i\epsilon\ln\epsilon}
\left (2+\left (-\pi +{1\over 36}\,mq-{2\over 3}\,{i}\kappa
\right )\epsilon\right )
\end{eqnarray*}

\begin{eqnarray*}
R^{\rm up}_{\ell m\omega}&=&
-45\,{\frac {{i}}{{z}^{2}}}-30\,{z}^{-1}+{15\over 2}\,{i}
+{5\over 8}\,{i}{z}^{2} \\
&&
+\epsilon\,\Biggl(\Bigl({\frac {45}{4}}\,mq
-45\,{i}\kappa+90\,{i}\Bigl (-{1\over 2}+{1\over 2}\,\kappa\Bigr )\Bigr){z}^{-3} \\
&&
+\Bigl(45\,{i}\pi +30+{15\over 2}\,\kappa-45\,\gamma
-{\frac {15}{8}}\,{i}mq\Bigr){z}^{-2}+\Bigl(30\,\pi -45\,{i} \\
&&
+{\frac {35}{24}}\,mq+30\,{i}\gamma-5\,{i}\kappa\Bigr){z}^{-1}\Biggr) \\
&&
+{\epsilon}^{2}\Biggl({\frac {135}{8}}\,\kappa\,mq-{
\frac {75}{2}}\,{i}{\kappa}^{2} \\
&&
+135\,{i}\kappa\,\Bigl (-{1\over 2}+{1\over 2}\,\kappa\Bigr )
-{\frac {135}{4}}\,\Bigl (-{1\over 2}+{1\over 2}\,\kappa\Bigr )mq
+{\frac {15}{8}}\,{i}{m}^{2}{q}^{2} \\
&&
-45\,{i}\Bigl (-\Bigl (-{1\over 2}+{1\over 2}\,\kappa\Bigr )^{2}
+2\,\Bigl (-1+\kappa\Bigr )\Bigl (
-{1\over 2}+{1\over 2}\,\kappa\Bigr )\Bigr )\Biggr){z}^{-4}\,.
\end{eqnarray*}

\noindent
(c) $\ell=4$

\begin{eqnarray*}
A^{\rm in}_{\ell m\omega}&&=
{1\over \omega}{1\over \kappa^6 \epsilon^6}
e^{{1\over 2}\,\pi i \,\left(\nu+3\right)} e^{i \epsilon \kappa} 
e^{-i\epsilon\ln\epsilon}
\Bigl\{
79380\,{\frac {{\kappa}^{2}}{\left (3\,\kappa+{i}mq\right )
\left (4\,\kappa+{i}mq\right )}}
\Bigr\}\,,
\end{eqnarray*}

\begin{eqnarray*}
C_{\ell m\omega}&&=
\left(\omega\over \epsilon\kappa\right)^{4}
e^{i\epsilon_{+}\ln\kappa}
\left (1+O(\epsilon) \right )\,,
\end{eqnarray*}

\begin{eqnarray*}
D_{\ell m\omega}=
{\omega}^{3}2^{i\epsilon} e^{-{\pi\over 2}i(\nu-1)}
e^{i\epsilon\ln\epsilon}
\left (2+O(\epsilon)\right )\,,
\end{eqnarray*}

\begin{eqnarray*}
R^{\rm up}_{\ell m\omega}=-{630 i\over z^3}\,.
\end{eqnarray*}

\section{The Energy Absorption Rate for Each Mode}
The energy absorption rate for each mode to $O(v^8)$ 
are given by 

\begin{eqnarray}
\eta_{2,2}&=&
-{1\over 4}\,q-{3\over 4}\,{q}^{3}+{v}^{2}\left (-{3\over 4}\,q
-{9\over 4}\,{q}^{3}\right )
\nonumber\\
&&
+\left({\frac {27}{4}}\,{q}^{2}+2\,q{\it B_2}+{\frac {15}{4}}\,{q}^{4}
+6\,{q}^{3}{\it B_2}+{13\over 2}\,\kappa\,{q}^{2}+3\,{q}^{4}\kappa
+{1\over 2}\,\kappa+{1\over 2}\right ){v}^{3}\nonumber\\
&&
+\left (-{\frac {199}{42}}\,q-{\frac {593}{42}}\,{q}^{3}
+{2\over 7}\,{q}^{5}\right ){v}^{4} \nonumber\\
&&
+\Bigl({\frac {721}{36}}\,{q}^{2}
+6\,q{\it B_2}+{\frac {127}{12}}\,{q}^{4}+18\,{q}^{3}{\it B_2}+{\frac {39}{2}}
\,\kappa\,{q}^{2}+9\,{q}^{4}\kappa \nonumber\\
&&
+{3\over 2}\,\kappa+{3\over 2}\Bigr){v}^{5} \nonumber\\
&&
+\Bigl(-{\frac {607076}{11025}}\,q-4\,{\it B_2}+{\frac {428}{105}}\,
\gamma\,q+{2\over 3}\,{\pi }^{2}q+{\frac {428}{105}}\,q\ln 2 \nonumber\\
&&
-4\,q{\it C_2}-12\,{q}^{3}{\it C_2}-36\,{q}^{4}{\it B_2}-56\,{q}^{2}{\it B_2}
+{\frac {428}{35}}\,{q}^{3}\gamma+{\frac {428}{35}}\,{q}^{3}\ln 2\nonumber\\
&&
+2\,{q}^{3}{\pi }^{2}
+{\frac {428}{105}}\,q\ln\kappa+{\frac {428}{105}}\,q{\it A_2}
+{\frac {428}{35}}\,{q}^{3}\ln\kappa+6\,{\frac {{q}^{7}}{\kappa}}
+{\frac {428}{35}}\,{q}^{3}{\it A_2}\nonumber\\
&&
-8\,q{{\it B_2}}^{2}-24\,{q}^{3}{{\it B_2}}^{2}
+{\frac {856}{105}}\,q\ln v+{\frac {856}{35}}\,{q}^{3}
\ln v-4\,{\frac {{\it B_2}}{\kappa}}-32\,{\frac {{q}^{3}}{\kappa}}
\nonumber\\
&&
-31\,{\frac {q}{\kappa}}+57\,{\frac {{q}^{5}}{\kappa}}-48\,{\frac {{q}^{2
}{\it B_2}}{\kappa}}+28\,{\frac {{q}^{4}{\it B_2}}{\kappa}}-24\,{\frac {
{q}^{3}{\it C_2}}{\kappa}}-8\,{\frac {q{\it C_2}}{\kappa}}
\nonumber\\
&&
+24\,{\frac {{
q}^{6}{\it B_2}}{\kappa}}-{\frac {11883052}{99225}}\,{q}^{3}+{\frac {
548}{27}}\,{q}^{5}\Bigr){v}^{6}
\nonumber\\
&&
+\Bigl({\frac {16747}{126}}\,{q}^{2}
+{\frac {15893}{189}}\,{q}^{4}-{\frac {155}{126}}\,{q}^{6}+123\,\kappa
\,{q}^{2}+{\frac {1142}{21}}\,{q}^{4}\kappa-{\frac {8}{7}}\,\kappa\,{q
}^{6}
\nonumber\\
&&
+{\frac {199}{21}}\,\kappa+{\frac {199}{21}}+{\frac {796}{21}}\,q
{\it B_2}+{\frac {2372}{21}}\,{q}^{3}{\it B_2}-{\frac {16}{7}}\,{q}^{5}{
\it B_2}\Bigr){v}^{7}
\nonumber\\
&&
+\Bigl(-{\frac {761349}{3920}}\,q-12\,{\it B_2}+
{\frac {3076}{105}}\,\gamma\,q+2\,{\pi }^{2}q+{\frac {4868}{105}}\,q
\ln 2-12\,q{\it C_2}
\nonumber\\
&&
-36\,{q}^{3}{\it C_2}-{\frac {308}{3}}\,{q}^{4}{\it B_2}
-{\frac {1496}{9}}\,{q}^{2}{\it B_2}+{\frac {3076}{35}}\,{q}^{3}
\gamma+{\frac {4868}{35}}\,{q}^{3}\ln 2+6\,{q}^{3}{\pi }^{2}
\nonumber\\
&&
+{\frac {428}{35}}\,q\ln\kappa+{\frac {428}{35}}\,q{\it A_2}
+{\frac {1284}{35}}\,{q}^{3}\ln\kappa
+{\frac {46}{3}}\,{\frac {{q}^{7}}{\kappa}}+{
\frac {1284}{35}}\,{q}^{3}{\it A_2}-24\,q{{\it B_2}}^{2}
\nonumber\\
&&
-72\,{q}^{3}{{\it B_2}}^{2}
+{\frac {872}{21}}\,q\ln v+{\frac {872}{7}}\,{q}^{3}\ln 
(v)-12\,{\frac {{\it B_2}}{\kappa}}-{\frac {272}{3}}\,{\frac {{q}^{3}}{
\kappa}}-{\frac {833}{9}}\,{\frac {q}{\kappa}}
\nonumber\\
&&
+{\frac {1511}{9}}\,{
\frac {{q}^{5}}{\kappa}}-144\,{\frac {{q}^{2}{\it B_2}}{\kappa}}+84\,{
\frac {{q}^{4}{\it B_2}}{\kappa}}-72\,{\frac {{q}^{3}{\it C_2}}{\kappa}}
-24\,{\frac {q{\it C_2}}{\kappa}}
\nonumber\\
&&
+72\,{\frac {{q}^{6}{\it B_2}}{\kappa}}
-{\frac {140529967}{317520}}\,{q}^{3}+{\frac {46465}{756}}\,{q}^{5}-{
\frac {191}{588}}\,{q}^{7}\Bigr){v}^{8}. 
\end{eqnarray}

\begin{eqnarray}
\eta_{2,1}&=&
{v}^{2}\left ({3\over 16}\,{q}^{3}-{1\over 4}\,q\right )
+\left (-{1\over 4}\,{q}^{4}+{1\over 3}\,{
q}^{2}\right ){v}^{3} \nonumber\\
&&
+\left ({1\over 12}\,{q}^{5}+{\frac {8}{9}}\,{q}^{3}-{4\over 3}
\,q\right ){v}^{4}\nonumber\\
&&
+\left (-{\frac {137}{48}}\,{q}^{4}-{3\over 4}\,{q}^{3}{
\it B_1}+{\frac {265}{72}}\,{q}^{2}+q{\it B_1}
-{3\over 4}\,{q}^{4}\kappa+{1\over 2}\,
\kappa+{\frac {7}{8}}\,\kappa\,{q}^{2}+{1\over 2}\right ){v}^{5} \nonumber\\
&&
+\left (-{\frac {382}{63}}\,q+{\frac {865}{756}}\,{q}^{3}+{\frac {2321}{1008}}\,
{q}^{5}-{2\over 3}\,\kappa\,q-{7\over 6}\,\kappa\,{q}^{3}+{q}^{5}\kappa+{q}^{4}{\it 
B_1}-{4\over 3}\,{q}^{2}{\it B_1}\right ){v}^{6} \nonumber\\
&&
+\Bigl({8\over 3}+{\frac {4643}{252}}
\,{q}^{2}-{\frac {32}{9}}\,{q}^{3}{\it B_1}-{\frac {1405}{3024}}\,{q}^{
6}-{\frac {65}{18}}\,{q}^{4}\kappa-{\frac {123269}{9072}}\,{q}^{4} \nonumber\\
&&
+{\frac {44}{9}}\,\kappa\,{q}^{2}-{1\over 3}\,{q}^{5}{\it B_1}
-{1\over 3}\,\kappa\,{q}^{6}
+{16\over 3}\,q{\it B_1}+{8\over 3}\,\kappa\Bigr){v}^{7} \nonumber\\
&&
+\Bigl(-{\frac {6292409
}{176400}}\,q-2\,{\it B_1}+{\frac {107}{105}}\,\gamma\,q
+{1\over 6}\,{\pi }^{2}q
+{\frac {107}{105}}\,q\ln 2-{\frac {107}{140}}\,{q}^{3}\gamma \nonumber\\
&&
-{\frac {107}{140}}\,{q}^{3}\ln 2-{1\over 8}\,{q}^{3}{\pi }^{2}+{\frac {107}{
105}}\,q\ln\kappa-{\frac {107}{140}}\,{q}^{3}\ln\kappa-{\frac {
55}{6}}\,{\frac {{q}^{7}}{\kappa}}+{\frac {214}{105}}\,q\ln v \nonumber\\
&&
-{\frac {107}{70}}\,{q}^{3}\ln v+{\frac {65}{12}}\,{\frac {{q}^{3}}{
\kappa}}-{\frac {245}{18}}\,{\frac {q}{\kappa}}+{\frac {625}{36}}\,{
\frac {{q}^{5}}{\kappa}}+{\frac {73}{6}}\,{q}^{4}{\it B_1}-q{\it C_1} \nonumber\\
&&
-{\frac {283}{18}}\,{q}^{2}{\it B_1}+{3\over 2}\,{q}^{3}{{\it B_1}}^{2}+{\frac {
107}{105}}\,q{\it A_1}-{\frac {107}{140}}\,{q}^{3}{\it A_1}-3\,{\frac {{
q}^{6}{\it B_1}}{\kappa}}+{13\over 2}\,{\frac {{q}^{4}{\it B_1}}{\kappa}} \nonumber\\
&&
-{3\over 2}\,{\frac {{q}^{2}{\it B_1}}{\kappa}}
-2\,{\frac {q{\it C_1}}{\kappa}}+{3\over 2}\,
{\frac {{q}^{3}{\it C_1}}{\kappa}}-2\,q{{\it B_1}}^{2}+{3\over 4}\,{q}^{3}{\it 
C_1}-2\,{\frac {{\it B_1}}{\kappa}} \nonumber\\
&&
+{\frac {381643}{6350400}}\,{q}^{3}+{
\frac {6439}{378}}\,{q}^{5}-{\frac {11}{168}}\,{q}^{7}\Bigr){v}^{8}. 
\end{eqnarray}

\begin{eqnarray}
\eta_{3,3}&=&
\left (-{\frac {75}{112}}\,{q}^{5}-{\frac {555}{896}}\,{q}^{3}-{\frac 
{15}{224}}\,q\right ){v}^{4}+\left (-{\frac {45}{112}}\,q-{\frac {1665
}{448}}\,{q}^{3}-{\frac {225}{56}}\,{q}^{5}\right ){v}^{6} \nonumber\\
&&
+\Bigl({\frac {225}{28}}\,{q}^{5}{\it B_3}+{\frac {375}{112}}\,{q}^{6}
+{\frac {15}{112}}+{\frac {1905}{448}}\,\kappa\,{q}^{2}
+{\frac {2055}{224}}\,{q}^{4}\kappa
+{\frac {15}{112}}\,\kappa \nonumber\\
&&
+{\frac {1665}{224}}\,{q}^{3}{\it B_3}
+{\frac {45}{56}}\,q{\it B_3}+{\frac {10995}{896}}\,{q}^{4}
+{\frac {2055}{448}}\,{q}^{2}\Bigr){v}^{7} \nonumber\\
&&
+\left (-{\frac {17697}{896}
}\,{q}^{3}-{\frac {18315}{896}}\,{q}^{5}+{\frac {125}{112}}\,{q}^{7}-{
\frac {481}{224}}\,q\right ){v}^{8}.
\end{eqnarray}

\begin{eqnarray}
\eta_{3,2}&=&
\left ({\frac {25}{189}}\,{q}^{5}-{\frac {5}{63}}\,q-{\frac {110}{567}
}\,{q}^{3}\right ){v}^{6} \nonumber\\
&&
+\left ({\frac {5}{42}}\,{q}^{2}+{\frac {55}{
189}}\,{q}^{4}-{\frac {25}{126}}\,{q}^{6}\right ){v}^{7} \nonumber\\
&&
+\left ({\frac {25}{336}}\,{q}^{7}-{\frac {40}{63}}\,q-{\frac {14485}{9072}}\,{
q}^{3}+{\frac {205}{216}}\,{q}^{5}\right ){v}^{8}. 
\end{eqnarray}

\begin{eqnarray}
\eta_{3,1}&=&
\left (-{\frac {1}{224}}\,q+{\frac {59}{8064}}\,{q}^{3}-{\frac {1}{336
}}\,{q}^{5}\right ){v}^{4} \nonumber\\
&&
+\left ({\frac {767}{12096}}\,{q}^{3}-{
\frac {13}{504}}\,{q}^{5}-{\frac {13}{336}}\,q\right ){v}^{6} \nonumber\\
&&
+\Bigl({
\frac {1}{84}}\,{q}^{5}{\it B_1}+{\frac {1}{56}}\,q{\it B_1}+{\frac {31}
{4032}}\,\kappa\,{q}^{2}-{\frac {55}{2016}}\,{q}^{4}\kappa+{\frac {1}{
84}}\,\kappa\,{q}^{6}+{\frac {109}{3024}}\,{q}^{6} \nonumber\\
&&
-{\frac {6395}{72576
}}\,{q}^{4}-{\frac {59}{2016}}\,{q}^{3}{\it B_1}+{\frac {1}{112}}+{
\frac {185}{4032}}\,{q}^{2}+{\frac {1}{112}}\,\kappa\Bigr){v}^{7} 
\nonumber\\
&&
+\left (-{\frac {1}{336}}\,{q}^{7}-{\frac {431}{2016}}\,q+{\frac {25105
}{72576}}\,{q}^{3}-{\frac {3271}{24192}}\,{q}^{5}\right ){v}^{8}.
\end{eqnarray}

\begin{eqnarray}
\eta_{4,4}&&=
-{\frac {5}{2268}}\,{v}^{8}q\left (9+7\,{q}^{2}\right )\left (3\,{q}^{
2}+1\right )\left (15\,{q}^{2}+1\right ).
\end{eqnarray}

\begin{eqnarray}
\eta_{4,2}&&=
-{\frac {5}{63504}}\,{v}^{8}q\left (5\,{q}^{2}-9\right )\left (3\,{q}^
{2}-4\right )\left (3\,{q}^{2}+1\right ).
\end{eqnarray}

\section{The Alternative Form of $(dE/dt)_{\rm H}$}
In this section, we describe the absorption rate $(dE/dt)_{\rm H}$ by means of 
$x$ $\equiv(M\Omega)^{1/3}$. Here we define 

$$
\left(dE\over dt \right)_{\rm Q}
={32\over 5}\left(\mu\over M\right)^2 x^{10}.
$$
Using the relation
$$
v=x\bigl(1+{1\over 3} q x^3+{2\over 9} q^2 x^6 +O(x^9)\bigr), 
$$
we have 
\begin{eqnarray*}
\left( {dE\over dt}\right)_{\rm H}&=&
\left(dE\over dt\right)_{\rm Q}\, x^5\, \Biggl[
-{1\over 4}\,q-{3\over 4}\,{q}^{3}
+\bigl (-q-{\frac {33}{16}}\,{q}^{3}\bigr ){x}^{2} \\
&&
+\bigl (2\,q{\it B_2}+{1\over 2}+{13\over 2}\,
\kappa\,{q}^{2}+{\frac {35}{6}}\,{q}^{2}-{1\over 4}\,{q}^{4}
+{1\over 2}\,\kappa \\
&&
+3\,{q}^{4}\kappa+6\,{q}^{3}{\it B_2}\bigr ){x}^{3} \\
&&
+\bigl (-{\frac {43}{7}}\,q-{\frac {17}{56}}\,{q}^{5}-{\frac {
4651}{336}}\,{q}^{3}\bigr ){x}^{4}\\
&&
+\bigl ({\frac {433}{24}}\,{q}^{2}-{\frac {95}{24}}\,{q}^{4}+2
-{3\over 4}\,{q}
^{3}{\it B_1}+2\,\kappa+{\frac {33}{4}}\,{q}^{4}\kappa+6\,q{\it B_2} \\
&&
+18\,{q}^{3}{\it B_2}+{\frac {163}{8}}\,\kappa\,{q}^{2}
+q{\it B_1}\bigr){x}^{5} \\
&&
+\bigl (-{\frac {2586329}{44100}}\,q-4\,{\it B_2}
-{\frac {1640747}{19600}}\,{q}^{3}+19\,{q}^{5}\kappa 
+{\frac {428}{105}}\,\gamma\,q 
+{2\over 3}\,{\pi }^{2}q \\
&&
+{\frac {428}{105}}\,q\ln (2)-4\,q{\it C_2}
-12\,{q}^{3}{\it C_2}-44\,{q}^{2}{\it B_2}+{\frac {428}{35}}\,{q}^{3}
\gamma+{\frac {428}{35}}\,{q}^{3}\ln (2) \\
&&
+2\,{q}^{3}{\pi }^{2} 
+{\frac {428}{105}}\,q\ln (\kappa)+{\frac {428}{105}}\,q{\it A_2}
+{\frac {428}{35}}\,{q}^{3}\ln (\kappa)+6\,{\frac {{q}^{7}}{\kappa}} \\
&&
+{\frac {428}{35}}\,{q}^{3}{\it A_2}-8\,q{{\it B_2}}^{2}
-24\,{q}^{3}{{\it B_2}}^{2}-4\,{\frac {{\it B_2}}{\kappa}}
-32\,{\frac {{q}^{3}}{\kappa}}-31\,{\frac {q}{\kappa}} \\
&&
+57\,{\frac {{q}^{5}}{\kappa}}+{q}^{4}{\it B_1}
-{4\over 3}\,{q}^{2}{\it B_1}+{7\over 3}\,\kappa\,q+{\frac {227}{6}}\,\kappa\,{q}^{3}
+{\frac {455}{16}}\,{q}^{5}\\
&&
-48\,{\frac {{q}^{2}{\it B_2}}{\kappa}}+28\,{\frac {{q}^{4}{\it B_2}}{\kappa}}
-24\,{\frac {{q}^{3}{\it C_2}}{\kappa}}-8\,{\frac {q{\it C_2}}{\kappa}} \\
&&
+24\,{\frac {{q}^{6}{\it B_2}}{\kappa}}
+{\frac {856}{105}}\,q\ln (x)+{\frac {856}{35}}\,{q}^{3}\ln (x)\bigr ){x}^{6} \\
&&
+\bigl ({\frac {19687}{168}}\,{q}^{2}
-{\frac {145}{336}}\,{q}^{6}-{\frac {4729}{1008}}\,{q}^{4}
+{\frac {899}{168}}\,q{\it B_1}
-{\frac {41}{28}}\,\kappa\,{q}^{6} \\
&&
+{\frac {45}{56}}\,q{\it B_3}
-{\frac {803}{224}}\,{q}^{3}{\it B_1}
+{\frac {1665}{224}}\,{q}^{3}{\it B_3}
+{\frac {86}{7}}+{\frac {719}{12}}\,{q}^{4}\kappa \\
&&
+{\frac {796}{21}}\,q{\it B_2}+{\frac {86}{7}}\,\kappa
-{\frac {16}{7}}\,{q}^{5}{\it B_2}+{\frac {225}{28}}\,{q}^{5}{\it B_3} \\
&&
-{\frac {9}{28}}\,{q}^{5}{\it B_1}+{\frac {22201}{168}}\,\kappa\,{q}^{2}
+{\frac {2372}{21}}\,{q}^{3}{\it B_2}\bigr ){x}^{7} \\
&&
\bigl (-{\frac {19366807}{88200}}\,q-12\,{\it B_2}-2\,{\it B_1}
-{\frac {2062220497}{6350400}}\,{q}^{3}-q{\it C_1}+55\,{q}^{5}\kappa \\
&&
+{\frac {1061}{35}}\,\gamma\,q+{\frac {13}{6}}\,{\pi }^{2}q
+{\frac {995}{21}}\,q\ln (2)-12\,q{\it C_2}-36\,{q}^{3}{\it C_2} \\
&&
+{\frac {52}{3}}\,{q}^{4}{\it B_2}-{\frac {1136}{9}}\,{q}^{2}{\it B_2}
+{\frac {12197}{140}}\,{q}^{3}\gamma
+{\frac {3873}{28}}\,{q}^{3}\ln (2) \\
&&
+{\frac {47}{8}}\,{q}^{3}{\pi }^{2}
+{\frac {1391}{105}}\,q\ln (\kappa)+{\frac {428}{35}}\,q{\it A_2}
+{\frac {5029}{140}}\,{q}^{3}\ln (\kappa) \\
&&
+{\frac {37}{6}}\,{\frac {{q}^{7}}{\kappa}}
+{\frac {1284}{35}}\,{q}^{3}{\it A_2}-24\,q{{\it B_2}}^{2}
-72\,{q}^{3}{{\it B_2}}^{2}-12\,{\frac {{\it B_2}}{\kappa}} \\
&&
-{\frac {341}{4}}\,{\frac {{q}^{3}}{\kappa}}
-{\frac {637}{6}}\,{\frac {q}{\kappa}}
+{\frac {741}{4}}\,{\frac {{q}^{5}}{\kappa}} \\
&&
+{\frac {43}{6}}\,{q}^{4}{\it B_1}
-{\frac {163}{18}}\,{q}^{2}{\it B_1}
+{3\over 2}\,{q}^{3}{{\it B_1}}^{2}+{\frac {107}{105}}\,q{\it A_1} \\
&&
-{\frac {107}{140}}\,{q}^{3}{\it A_1}-2\,q{{\it B_1}}^{2}
+{3\over 4}\,{q}^{3}{\it C_1}-2\,{\frac {{\it B_1}}{\kappa}}
+{\frac {40}{3}}\,\kappa\,q+{\frac {815}{6}}\,\kappa\,{q}^{3} \\
&&
+{\frac {1265}{18}}\,{q}^{5}+{\frac {25}{252}}\,{q}^{7}
-144\,{\frac {{q}^{2}{\it B_2}}{\kappa}}+84\,{\frac {{q}^{4}{\it B_2}}{\kappa}}
-72\,{\frac {{q}^{3}{\it C_2}}{\kappa}}-24\,{\frac {q{\it C_2}}{\kappa}} \\
&&
+72\,{\frac {{q}^{6}{\it B_2}}{\kappa}}-3\,{\frac {{q}^{6}{\it B_1}}{\kappa}}
+{13\over 2}\,{\frac {{q}^{4}{\it B_1}}{\kappa}} \\
&&
-{3\over 2}\,{\frac {{q}^{2}{\it B_1}}{\kappa}}-2\,{\frac {q{\it C_1}}{\kappa}}
+{3\over 2}\,{\frac {{q}^{3}{\it C_1}}{\kappa}}+{\frac {4574}{105}}\,q\ln (x)
+{\frac {8613}{70}}\,{q}^{3}\ln (x)\bigr ){x}^{8} \Biggr].
\end{eqnarray*}


\end{document}